\documentclass[aps,prd,onecolumn,showpacs,superscriptaddress,preprintnumbers,floatfix,reprint,nofootinbib]{revtex4-1}
\usepackage{graphicx}
\usepackage{amsmath}
\usepackage{bm}
\usepackage{yhmath}
\usepackage[colorlinks,citecolor=blue,urlcolor=blue,linkcolor=blue]{hyperref}
\usepackage{subfigure}
\usepackage{color}
\usepackage{cases}
\usepackage{subfigure}
\usepackage{dcolumn,booktabs,bm}
\usepackage{slashed}
\usepackage{amsfonts,amssymb,stmaryrd,latexsym,amsmath}
\usepackage{textcomp}
\usepackage{multirow}
\usepackage{cancel}
\usepackage{diagbox}
\usepackage{array}
\usepackage{url}

\renewcommand{\arraystretch}{1.8}
\allowdisplaybreaks

\begin{document}

\title{$S$-wave fully charmed tetraquark resonant states}
\author{Guang-Juan Wang}\email{wgj@pku.edu.cn}
\affiliation{Advanced Science Research Center, Japan Atomic Energy
Agency, Tokai, Ibaraki, 319-1195, Japan}

\author{Qi Meng}\email{qimeng@nju.edu.cn}
\affiliation{Department of Physics, Nanjing University, Nanjing 210093, P.R. China}

\author{Makoto Oka}\email{oka@post.j-parc.jp}
\affiliation{Advanced Science Research Center, Japan Atomic Energy
Agency, Tokai, Ibaraki, 319-1195, Japan} \affiliation{Nishina Center
for Accelerator-Based Science, RIKEN, Wako 351-0198, Japan}

\begin{abstract}
We calculate the mass spectrum of the $S$-wave fully-charmed tetraquark resonant states  $cc\bar c\bar c $ in the nonrelativistic quark model, which successfully describes the charmonium spectrum. 
 The four-body system is solved with the Gaussian expansion method. The complex scaling technique is used to identify the genuine resonances. With the nonrelativistic quark model, our results show the existence of two  $cc\bar c\bar c$ resonances in each of the $J^{PC}=$ $0^{++}$, $1^{+-}$ and $2^{++}$  sectors, respectively.  In the $S$-wave sector, no resonance is found at the energy region of the $X(6200)$ and $X(6600)$  states. The lower  $0^{++}$ and $2^{++}$ resonances are located around  $100$ MeV higher than the $X(6900)$ state observed in experiments but have the decay widths consistent with the experiment. The higher $0^{++}$  and $2^{++}$ resonances are found at around $7.2$ GeV with the widths of $60.6$ MeV and $91.2$ MeV, respectively, and they may be good candidates for the $X(7200)$ state. 
\end{abstract}
\maketitle

\section{Introduction}\label{intro}
Since 2003, the experiments reported dozens of the exotic states at the odd of the quark model predictions,  which are the good multiquark candidates of the tetraquark state $qq\bar q\bar q$ and pentaquark state $qqqq\bar q$.
Compared with the conventional $q\bar q$ mesons and $qqq$ baryons, the multiquark states (the number of inner quarks $N\geq 4$) have much richer confinement mechanisms to form the color-singlet hadrons.
Correspondingly, many theoretical interpretations of the inner structures have been proposed: hadronic molecule, compact tetraquark state, hadron-heavy quarkonium, and so on ( For more details, refer to
Refs.~\cite{Brambilla:2019esw,Chen:2016qju,Guo:2017jvc,Esposito:2016noz,Hosaka:2016pey,Ali:2017jda,Liu:2019zoy,Lebed:2016hpi,Meng:2022ozq}.). 
Up to now, there does not exist a multiquark state with well established inner structure.  The inner dynamics are still mysterious.

The fully heavy tetraquark state $QQ\bar Q \bar Q$ ($Q=b,c$) provides a clear environment to investigate the quark confinement dynamics in the multiquark system. Without a mechanism for creating light quarks, the coupled channel effect between the heavy quarkonium core $Q\bar Q$ and the tetraquark state $Q\bar Q q\bar q$ is ruled out.   In the past years, the tetraquark $bb\bar b\bar b$ states  were searched  in the $\Upsilon(1S)\mu^+\mu^-$ channel by CMS~\cite{Khachatryan:2016ydm,CMS:2020qwa} and LHCb collaborations~\cite{Aaij:2018zrb}.  No significant excess for a narrow resonance was seen.   Recently there are great progress in the search for the  $cc\bar c\bar c$  states. In 2020, the LHCb reported the observation of a  broad structure with the mass ranging in $(6.2, 6.8)$ GeV and a resonance $X(6900)$  in the  di-$J/\psi$ channel~\cite{Aaij:2020fnh}.  The $X(6900)$ was confirmed in the di-$J/\psi$ channel in CMS collaboration \cite{CMS}, and di-$J/\psi$ as well as $J/\psi\psi(2S)$ channels in ATLAS collaboration \cite{ATLAS}, respectively. In addition, there are new resonances reported, the $X(6600)$ and $X(7200)$ in CMS \cite{CMS}, and the $X(6200)$, $X(6600)$ as well as  $X(7200)$ in ATLAS \cite{ATLAS}. The parameters of the resonances and their observable channels are summarized in Table \ref{tab:ccccexperiments}.  The minimum constituents of these states are $cc\bar c\bar c$, which may be good candidates to pin down the inner dynamics of the multiquark states. 

The theoretical studies of the fully heavy quark states started long time ago before the experimental results~\cite{Iwasaki:1975pv,Chao:1980dv,Ader:1981db,Zouzou:1986qh,Heller:1986bt,SilvestreBrac:1992mv,SilvestreBrac:1993ry}. Recent experimental progresses have inspired  intensive theoretical investigations \cite{Albuquerque:2020hio,liu:2020eha,Jin:2020jfc,Lu:2020cns,Giron:2020wpx,Dosch:2020hqm,Yang:2020wkh,Huang:2020dci,Hughes:2021xei,Faustov:2021hjs,Liang:2021fzr,Li:2021ygk,Guo:2020pvt,Dong:2020nwy,Wan:2020fsk,
    Zhu:2020snb,Feng:2020riv,Wang:2020gmd,Maciula:2020wri,Feng:2020qee, Goncalves:2021ytq,Huang:2021vtb,Mutuk:2021hmi,Lu:2020cns,Chen:2020xwe,Becchi:2020uvq,An:2020jix,Bedolla:2019zwg,Cao:2020gul,Chao:2020dml,Faustov:2020qfm,Gong:2020bmg,Gordillo:2020sgc,Karliner:2020dta,Ke:2021iyh,liu:2020eha,Lucha:2021mwx,Ma:2020kwb,Richard:2020hdw,Sonnenschein:2020nwn,Wang:2020dlo,Wang:2020wrp,Wang:2021xao,Weng:2020jao,Yang:2021hrb,Yang:2021zrc,Zhang:2020xtb,Zhao:2020cfi,Zhao:2020nwy,Zhu:2020xni,Wang:2021taf,Dong:2021lkh,Wang:2022jmb,Chen:2022sbf,Wu:2022qwd,Zhuang:2021pci} (More discussions are referred to the reviews \cite{Mai:2022eur,Chen:2022asf}.). The interpretations of their natures were discussed in different models, compact tetraquark states~\cite{Albuquerque:2020hio,liu:2020eha,Jin:2020jfc,Lu:2020cns,Giron:2020wpx,Dosch:2020hqm,Yang:2020wkh,Huang:2020dci,Hughes:2021xei,Faustov:2021hjs,Liang:2021fzr,Li:2021ygk,Liu:2021rtn,Zhou:2022xpd,Asadi:2021ids,Liu:2021rtn,Wang:2021taf,Yang:2021hrb,Ke:2021iyh,Wang:2021kfv,liu:2020eha},  the dynamical rescattering mechanism of double-charmonium channels~\cite{Guo:2020pvt,Dong:2020nwy,Gong:2022hgd,Wang:2020wrp,Gong:2020bmg,Guo:2020pvt,Liang:2022rew,Wang:2022jmb}, $c\bar c$ hybrid~\cite{Wan:2020fsk}, and a Higgs-like boson~\cite{Zhu:2020snb}. Since the heavy quarks cannot exchange the light mesons, the compact tetraquark interpretations of the $QQ\bar Q\bar Q$ states are natural and popular. However, one struggling problem is the predictions of many more additional tetraquark states than the experimental observations.  In our previous works \cite{Wang:2019rdo, Wang:2021kfv}, we have calculated the mass spectrum of the $S$-wave and $P$-wave as well as the radially excited fully heavy tetraquark states in the constituent quark model. The results show that the three lowest $cc\bar c\bar c$ states are located in the mass region $(6.5,6.7, 6.9)$ GeV, respectively.  The obtained states are much more than the experimental observations. As discussed in these studies, the redundancy states may have two origins. On one hand, the finite number of the bases may lead to the discrete eigenvalues of the scattering states, which may be misidentified as the resonance. On the other hand, the multiquark states may decompose into the scattering states and may have large decay widths given the strong coupling effect, which cannot be observed in experiments. In this work, we propose to identify the resonant tetraquark states $cc\bar c\bar c$ with the complex scaling method.

The paper is arranged as follows. We give an introduction of the formulation in Sec. \ref{sec1}. We present the Hamiltonian in Sec. \ref{subsec1}, the four-body wave functions in Sec. \ref{subsec2}, and the complex scaling method in Sec. \ref{subsec3}, respectively. In Sec. \ref{sec2}. we discuss our results. At last, a summary is given in Sec. \ref{sec3}.
   \begin{table*}
 \renewcommand\arraystretch{1.5}
 \caption{The masses and decay widths of the tetraquark states reported in experiments. The results from LHCb~\cite{Aaij:2018zrb} and CMS~\cite{CMS}  are given in unit of MeV, while those in ATLAS~\cite{ATLAS}  are given in GeV.}
 \label{tab:ccccexperiments}
 \centering
 \setlength{\tabcolsep}{2.3mm}
\begin{tabular}{ccccc}
\toprule[1pt] 
 &  & $M$ & $\Gamma$ & Observable channels\tabularnewline
\midrule[1pt] 
LHCb model I~\cite{Aaij:2018zrb} & \multirow{2}{*}{$X(6900)$} & ${6905\pm11\pm7}$ & ${80\pm19\pm33}$ & \multirow{2}{*}{di-$J/\psi$}\tabularnewline
LHCb model II~\cite{Aaij:2018zrb}  &  & ${6886\pm11\pm11\mathrm{}}$ & ${168\pm33\pm69}$ & \tabularnewline
\hline 
\multirow{3}{*}{CMS~\cite{CMS} } & ${X(6600)}$ & ${6552\pm10\pm12\mathrm{}}$ & ${124\pm29\pm34\mathrm{}}$ & \multirow{3}{*}{di-$J/\psi$}\tabularnewline
 
  & $X(6900)$ & ${6927\pm9\pm5\mathrm{}}$ & ${122\pm22\pm19}$ & \tabularnewline
  & ${X(7200)}$ & $7{287\pm19\pm5}$ & ${95\pm46\pm20}$ & \tabularnewline
\hline 
\multirow{5}{*}{ATLAS~\cite{ATLAS} } & ${X(6200)}$ & ${6.22\pm0.05_{-0.05}^{+0.04}}$ & ${0.31\pm0.12_{-0.08}^{+0.07}}$ & \multirow{3}{*}{di-$J/\psi$}\tabularnewline
  & ${X(6600)}$ & ${6.62\pm0.03_{-0.01}^{+0.02}\mathrm{}}$ & ${0.31\pm0.09_{-0.11}^{+0.06}}$ & \tabularnewline
  \cline{2-4}
  & \multirow{2}{*}{${X(6900)}$${}$} & ${6.87\pm0.03_{-0.01}^{+0.06}}$ & ${0.12\pm0.04_{-0.01}^{+0.03}}$ & \tabularnewline
    \cline{3-5}
  &  & ${6.78\pm0.36_{-0.54}^{+0.35}\mathrm{}}$ & ${0.39\pm0.11_{-0.07}^{+0.11}}$ & \multirow{2}{*}{$J/\psi\psi(2S)$}\tabularnewline
    \cline{2-4}
  & ${X(7200)}$ & ${7.22\pm0.03_{-0.03}^{+0.02}}$ & ${0.10_{-0.07-0.05}^{+0.13+0.06}}$ & \tabularnewline
\bottomrule[1pt] 
\end{tabular}
\end{table*}

\section{Formulation} \label{sec1}
\subsection{Hamiltonian}\label{subsec1}
We introduce the nonrelativistic quark model for the fully heavy tetraquark $QQ\bar Q \bar Q$ system. The Hamiltonian is given by  
\begin{eqnarray}
H & =&\sum_{i=1}^{4}\frac{\mathbf{p}_{i}^{2}}{2m_{i}}+\sum_{i}m_{i}+\sum^4_{i<j=1}\frac{\mathbf{\lambda}_i}{2}\frac{\mathbf{\lambda}_j}{2}V_{ij},
\end{eqnarray}
where $\mathbf{p}_i$ and $m_i$ are the momentum in the c.m.s and mass of the
$i$th (anti)quark, respectively.  ${\mathbf \lambda}_i$ is the color matrix.  $V_{ij}$ is the potential between the $i$th and $j$th (anti-)quarks. In this work, we use the quark model proposed in Ref. \cite{Barnes:2005pb} (labeled as BGS in the following) to study the $cc\bar c\bar c$ system. The potential is given by
\begin{eqnarray}
V_{ij}&=&\frac{\alpha_{s}}{r_{ij}}-\frac{3}{4}br_{ij}-\frac{8\pi\alpha_{s}}{3m_im_j}\left(\frac{\sigma}{\sqrt{\pi}}\right)^{3}e^{-\sigma^{2}r_{ij}^{2}}\mathbf{s}_{i}\cdot\mathbf{s}_{j}. \nonumber\\
\end{eqnarray}
 It consists of the color-Coulomb, the linear confinement, and the hyperfine interactions. The charmonium spectrum determines all the parameters as listed in Table \ref{tab:par}. With the model, we can calculate the mass spectrum of the low-lying charmonium spectrum with the Gaussian expansion method and summarize the results in Table \ref{tab:meson}.   

\begin{table}
 \renewcommand\arraystretch{1.5}
 \caption{The values of parameters in the BGS quark model~\cite{Barnes:2005pb}. }\label{tab:par}
 \centering
 \setlength{\tabcolsep}{2.3mm}
\begin{tabular}{cccccccc}
\toprule[1pt]
\multirow{2}{*}{} \multirow{2}{*}{} $m_{c}$~{[}GeV{]} & $\alpha_s$& $b~[\text{GeV}^{2}]$ & $\sigma$~{[}GeV{]}&
   \tabularnewline
  $1.4794$ &  $0.5461$  &   $0.1425$ & $1.0946$ &   \tabularnewline
\bottomrule[1pt]
\end{tabular}
\end{table}

\begin{table}
 \renewcommand\arraystretch{1.5}
 \caption{The charmonium mass spectrum (in the unit of MeV) obtained with the Gaussian expansion method in the BGS quark model \cite{Barnes:2005pb}  compared with their experimental values (EXP)~\cite{Zyla:2020zbs}. }
 \label{tab:meson}
 \centering
 \setlength{\tabcolsep}{2.3mm}
\begin{tabular}{cccccc}
%\multicolumn{4}{c}{Mass spectrum (MeV)}\tabularnewline 
\toprule[1pt] 
 $^{2S+1}L_{J}$ & $c\bar c$ & EXP  &
BGS\tabularnewline 
\midrule[1pt] 
 $^{1}S_{0}$
& $\eta_{c}$ & $2983.9$ & $2982$\tabularnewline

 $^{3}S_{1}$ & $J/\psi$ & $3096.9$ & 
$3090$ \tabularnewline

$^{3}P_{0}$ & $\chi_{c0}$ & $3414.7$ & $3424$ \tabularnewline

$^{3}P_{1}$ & $\chi_{c1}$ & $3510.7$ & $3505$\tabularnewline

$^{1}P_{1}$ & $h_{c}(1P)$ & $3525.4$ & $3516$\tabularnewline

 $^{3}P_{2}$ & $\chi_{c2}$ & $3556.2$ & $3556$\tabularnewline

  $^{1}S_{0}$ & $\eta_{c}(2S)$ & $3637.5$ & $3630$ \tabularnewline

  $^{3}S_{1}$ & $\psi(2S)$ & $3686.1$ & $3672$ \tabularnewline

 $^{3}S_{1}$ & $\psi(3S)$ & $4039.0$  & $4072$ \tabularnewline

  $^{3}S_{1}$ & $\psi(4S)$ & $4421.0$ & $4420$ \tabularnewline

\bottomrule[1pt]
\end{tabular}
\end{table}

\subsection{The four-body calculation method}\label{subsec2}
\begin{figure}
\centering
\includegraphics[width=0.5\textwidth]{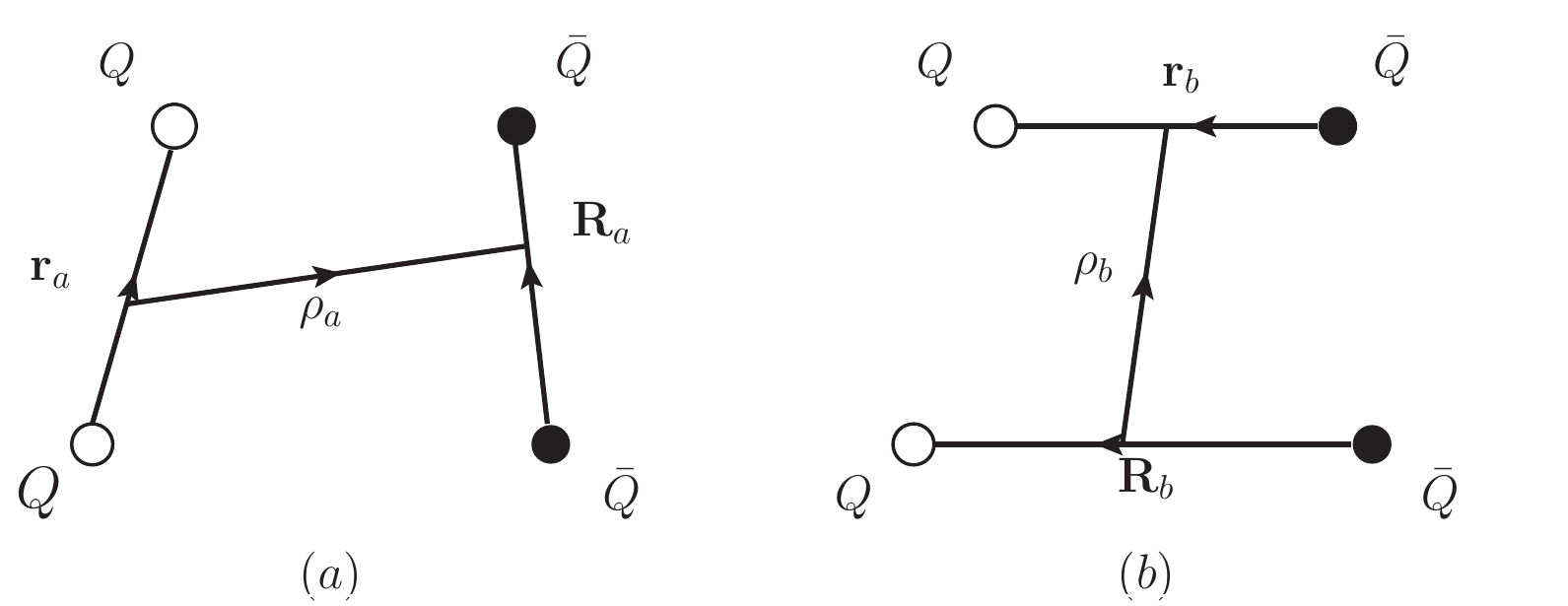}
\caption{(a): The diquark-antidiquark ($QQ$-$\bar Q\bar Q $) (b) Meson-meson ($Q\bar Q$-$Q\bar Q$) configurations in the four-quark system.} \label{fig:jac}
\end{figure}

For a $Q_1Q_2\bar Q_3\bar Q_4$ tetraquark state, there are two possible configurations, the diquark-antidiquark (a) and the meson-meson (b) ones as shown in Fig. \ref{fig:jac}. With the Gaussian expansion method, the wave function of the tetraquark state $\psi_{JM}$ is given by 
\begin{eqnarray}
\begin{aligned}
\psi_{JM}=&\sum_{C=a,b}\sum_{\alpha}A_{12}A_{34}\sum_\alpha \mathcal B^{(C)}_{\alpha}\chi^{(C)}_{C} \\
\times&\left[\left[\text{\ensuremath{\phi}}_{nl}(\mathbf {r}_{C})\otimes\text{\ensuremath{\phi}}_{NL}(\mathbf{R}_{C})\otimes\phi_{\nu\lambda}(\mathbf{\rho}_{C})\right]_{J_L}\otimes \chi^{(C)}_S\right]^{JM},
\end{aligned}
\end{eqnarray}
 where the  $\mathbf {r}_C$, $\mathbf{R}_C$, and $\mathbf{\rho}_C$ are three independent Jacobi coordinates as shown in Fig. \ref{fig:jac}. The superscript $C$ denotes the  $a$ and $b$ color configurations as illustrated in Fig. \ref{fig:jac}.  The $J_L$ and $S$ are the orbital angular momentum and the intrinsic spin of the tetraquark system, respectively. They combine to form the total angular momentum $J$. $A_{12}$/$A_{34}$ is the antisymmetric operator of the two quarks/antiquarks. $  \mathcal B_\alpha^{(C=a,b)}$ is the expanding coefficient of the $\alpha$th basis, which has  a set of  quantum number $\{nl,NL,\nu \lambda, J_{L},S\}$ forming the total spin-parity $J^P$. The $\chi^{(C)}_S$ and $\chi^{(^{C})}_C$  are the total spin and color wave-functions in the color configuration $C$ ($C$=$a$, $b$). Constrained by the Fermi statistic, the possible spin-color wave functions of the $QQ\bar Q \bar Q$ tetraquark state are given by 
 \begin{eqnarray}
&&\chi^{0^{++}}_{a,1}=[\{QQ\}_{\bar{3}_{c}}^{s=1}\{\bar Q\bar Q\}_{3_{c}}^{s=1}]^0, \label{eq:01} \\
&&\chi^{0^{++}}_{a,2}=[\{QQ\}_{6_{c}}^{s=0}\{\bar Q \bar Q \}_{\bar 6_{c}}^{s=0}]^0,\label{eq:02} \\
&&\chi^{0^{++}}_{b,1}=[\{Q \bar Q\}_{1_{c}}^{s=1}\{Q\bar Q\}_{1_{c}}^{s=1}]^0,\label{eq:03} \\
&&\chi^{0^{++}}_{b,2}=[\{Q \bar Q\}_{1_{c}}^{s=0}\{ Q \bar Q \}_{1_{c}}^{s=0}]^0,\label{eq:04}\\
&&\chi^{1^{+-}}_{a,1}=[\{QQ\}_{\bar{3}_{c}}^{s=1}\{\bar Q\bar Q\}_{3_{c}}^{s=1}]^1,\label{eq:11}\\
&&\chi^{1^{+-}}_{b,1}=[\{Q \bar Q\}_{1_{c}}^{s=1}\{Q\bar Q\}_{1_{c}}^{s=0}]^1,\label{eq:12}\\
&&\chi^{2^{++}}_{a,1}=[\{QQ\}_{\bar{3}_{c}}^{s=1}\{\bar Q\bar Q\}_{3_{c}}^{s=1}]^2,\label{eq:21}\\
&&\chi^{2^{++}}_{b,1}=[\{Q \bar Q\}_{1_{c}}^{s=1}\{Q\bar Q\}_{1_{c}}^{s=1}]^2.\label{eq:22}
 \end{eqnarray}
 where the superscript denotes the spin-parity quantum numbers.  The $\phi_{nl}(\mathbf{r}_C)$/ $\phi_{NL}(\mathbf{R}_C)$/ $\phi_{\nu\lambda}(\mathbf{\rho}_C)$ is the spatial wave function with  $n/N/\nu$ and $l/L/\lambda$  the radial quantum number and orbital  angular momentum, respectively. As an example, the $\phi_{nl}(\mathbf{r}_C)$ reads,
  \begin{eqnarray}
 &&\phi_{nlm}(\mathbf{r}_C)=N_{nl}r^{l}e^{-(r/r_n)^{2}}Y_{lm}(\hat{\mathbf{r}}_{C}),\\
 &&  r_{n}=r_{1}a^{n-1},\\
&&	a=(\frac{r_{n_{max}}}{r_{1}})^{\frac{1}{n_{max}-1}}, \,\,\,(n=1,...,n_{max}),
  \end{eqnarray}
The spatial wave functions $\phi_{NL}(\mathbf{R}_C)$ and $\phi_{\nu\lambda}(\mathbf{\rho}_C)$ are similar.

With the Hamiltonian and the wave functions, we can solve the Schr\"{o}dinger equation to obtain the eigenvalues. However, one should note that all the eigenvalues including those for the continuum states are discrete since we calculate the mass spectrum in a finite number of bases, which is equivalent to the finite volume.   To identify the resonance we apply the complex scaling method \cite{Aoyama:2006,Myo:2014ypa,moiseyev1998quantum}. 

\subsection{Complex scaling method} \label{subsec3}
In the complex scaling method, all the relative coordinate $\boldsymbol {r}$ and the momentum $\boldsymbol q$ will be scaled as follows,  
  \begin{eqnarray}
\boldsymbol{r} \rightarrow \boldsymbol{r} e^{i \theta}, \quad \boldsymbol{q} \rightarrow \boldsymbol{q} e^{-i \theta}.
  \end{eqnarray}
  where $\theta$ is a positive angle. Correspondingly, the Schr\"{o}dinger equation reads 
    \begin{eqnarray}
    H_{\theta} \Phi_{\theta}=E_{\theta} \Phi_{\theta},
      \end{eqnarray}
 with     
     \begin{eqnarray}
&& H_{\theta}={H}\left(\boldsymbol{r}  e^{i \theta}, \boldsymbol{q} e^{-i \theta}\right) \nonumber \\
 &&=\sum_{i=1}^{4}\frac{\mathbf{p}_{i}^{2}}{2m_{i}}e^{-i2\theta}+\sum_{i}m_{i}+\sum^4_{i<j=1}\frac{\mathbf{\lambda}_i}{2}\frac{\mathbf{\lambda}_j}{2}V_{ij}(\boldsymbol{r}_{ij}  e^{i \theta}).
   \end{eqnarray}
   Since we have expanded the states with the Gaussian bases, all the eigenvalues are discretized and complex. By applying the transformation,  the scattering states will rotate with $2\theta$ along the continuum line.  For the pole with $E_{\text {pole}}=E_{\text {resonance }}=E-i \frac{\Gamma}{2} $ ($E$ and $\Gamma$ being the energy and the decay width of the resonance), when the $\theta$ satisfies $\tan (2 \theta)>\Gamma /\left(2 E\right)$, the resonant wave functions will converge at $r\rightarrow \infty$ and can be expanded with the Gaussian wave functions similar to the bound states.    
The bound and resonant states will stay stable and do not move with the changing of $\theta$.  More technical details are referred to Refs. \cite{Aoyama:2006,Myo:2014ypa,moiseyev1998quantum}.

\section{Results} \label{sec2}
\begin{figure*}
\centering
\includegraphics[width=0.8\textwidth]{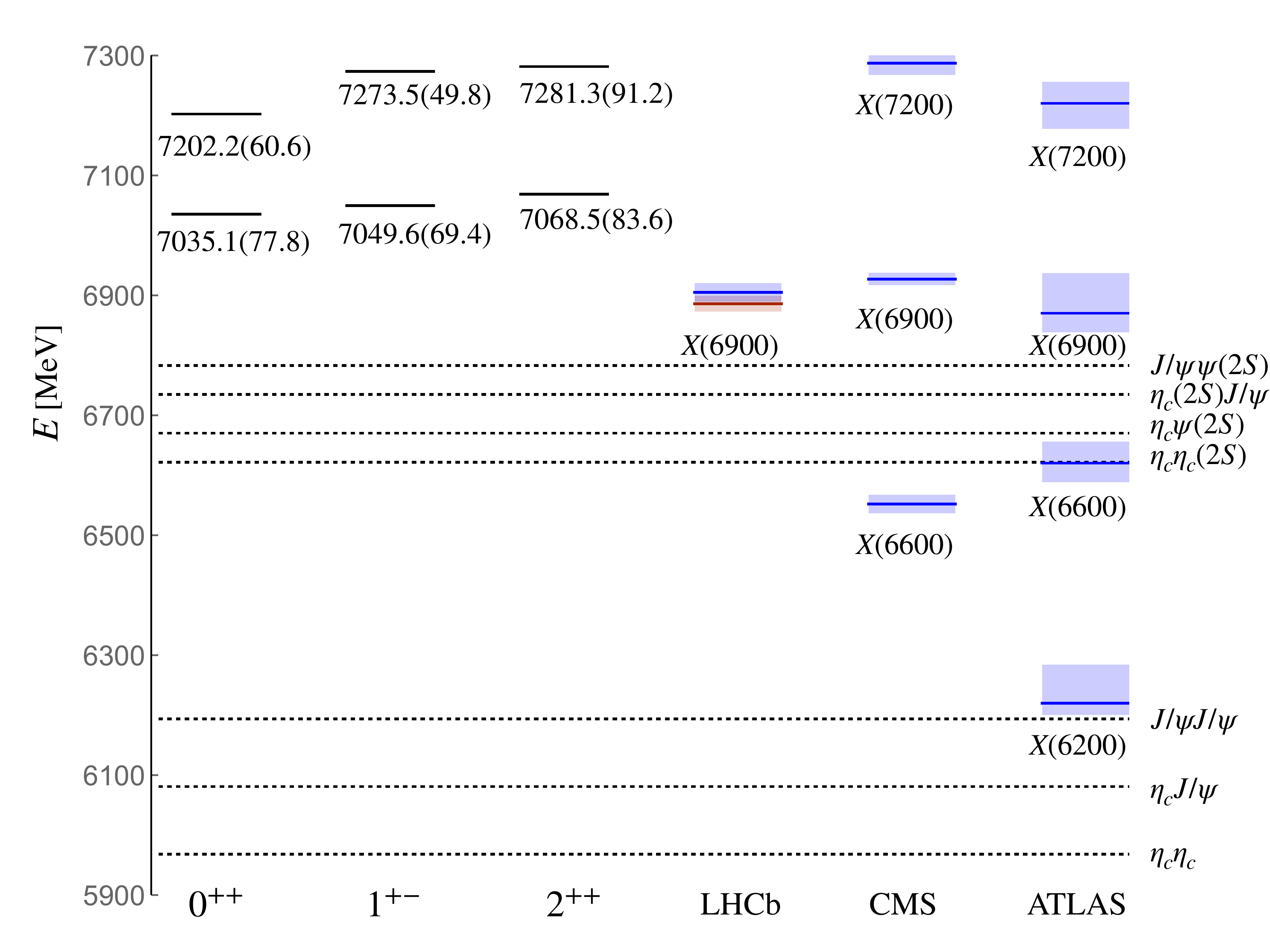}
\caption{The masses and widths (in the brackets) of $S$-wave $cc\bar c\bar c$ resonances obtained in the complex scaling method in the unit of MeV. The colored lines represent the central masses in LHCb~\cite{Aaij:2018zrb}, CMS~\cite{CMS}, and ATLAS~\cite{ATLAS}. The dashed area denotes the uncertainties. The red color for $X(6900)$ refers to the second fitting results obtained at LHCb. For the $X(6900)$ state in ATLAS, we choose the mass and decay width obtained in the di-$J/\psi$ channel instead of those in the $J/\psi\psi(2S)$ channel due to the large uncertainty. }  \label{fig:ccccBGS}
\end{figure*}

We use the quark model to calculate the $cc\bar c\bar c$ spectrums and summarize the results in Table \ref{tab:ccccBGS}. The comparisons between the theoretical and experimental results are displayed in Fig. \ref{fig:ccccBGS}. 
There are two resonances in the $J^{PC}=0^{++}$, $J^P=1^{+-}$ and $J^P=2^{++}$ sectors, respectively.  As shown in Fig. \ref{fig:ccccBGS}, all these six states are above the lowest dicharmonium channels, di-$\eta_c$, di-$J/\psi$ or $J/\psi\eta_c$. No bound states are obtained.

\begin{table}
 \renewcommand\arraystretch{1.5}
 \caption{The masses and half decay widths  $E_r-i\frac{\Gamma}{2}$ ( in unit of MeV) of the $cc\bar c \bar c$ resonances obtained with the BGS quark model~ \cite{Barnes:2005pb}. }
 \label{tab:ccccBGS}
 \centering
 \setlength{\tabcolsep}{2.3mm}
\begin{tabular}{cccc}
\toprule[1pt] 
%& \multicolumn{2}{c}{$E_r-i\frac{\Gamma}{2}$}\tabularnewline
$J^{PC}$  & 1st & 2nd \tabularnewline
 \midrule[1pt]  
$0^{++}$ & $7035.1 -i~38.9$   & $7202.2 - i~30.3$ \tabularnewline
 
$1^{+-}$ & $7049.6 - i~34.7$ & $7273.5 - i~24.9$\tabularnewline

$2^{++}$ & $7068.5 - i~41.8 $ & $7281.3 -i~45.6$  \tabularnewline
 
\bottomrule[1pt] 
\end{tabular}
\end{table}

\begin{figure*}
\centering \subfigure[]{
\begin{minipage}[t]{0.48\linewidth}
\centering
\includegraphics[width=1\textwidth]{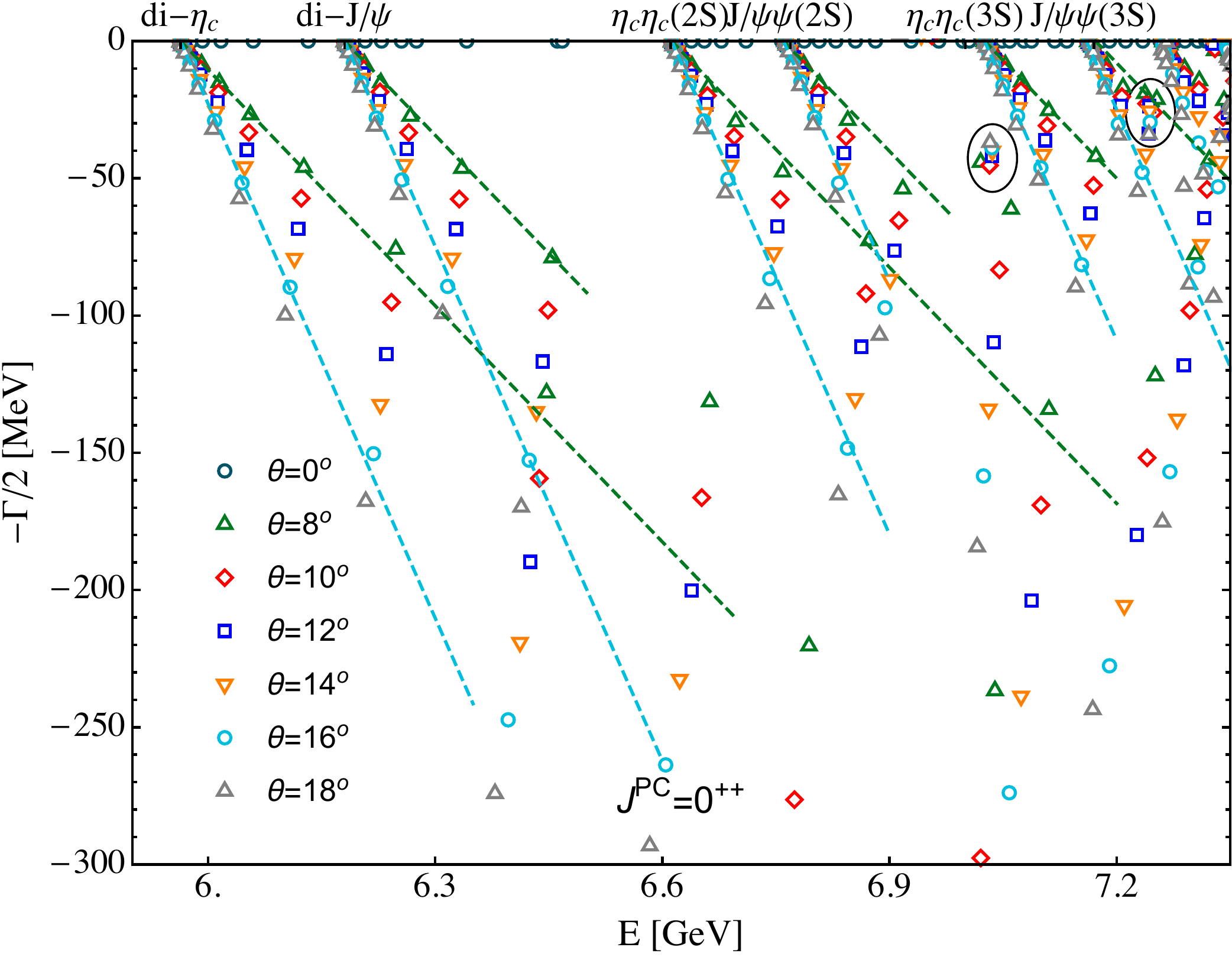}

\end{minipage}%
}%
\subfigure[]{
\begin{minipage}[t]{0.49\linewidth}
\centering
\includegraphics[width=1\textwidth]{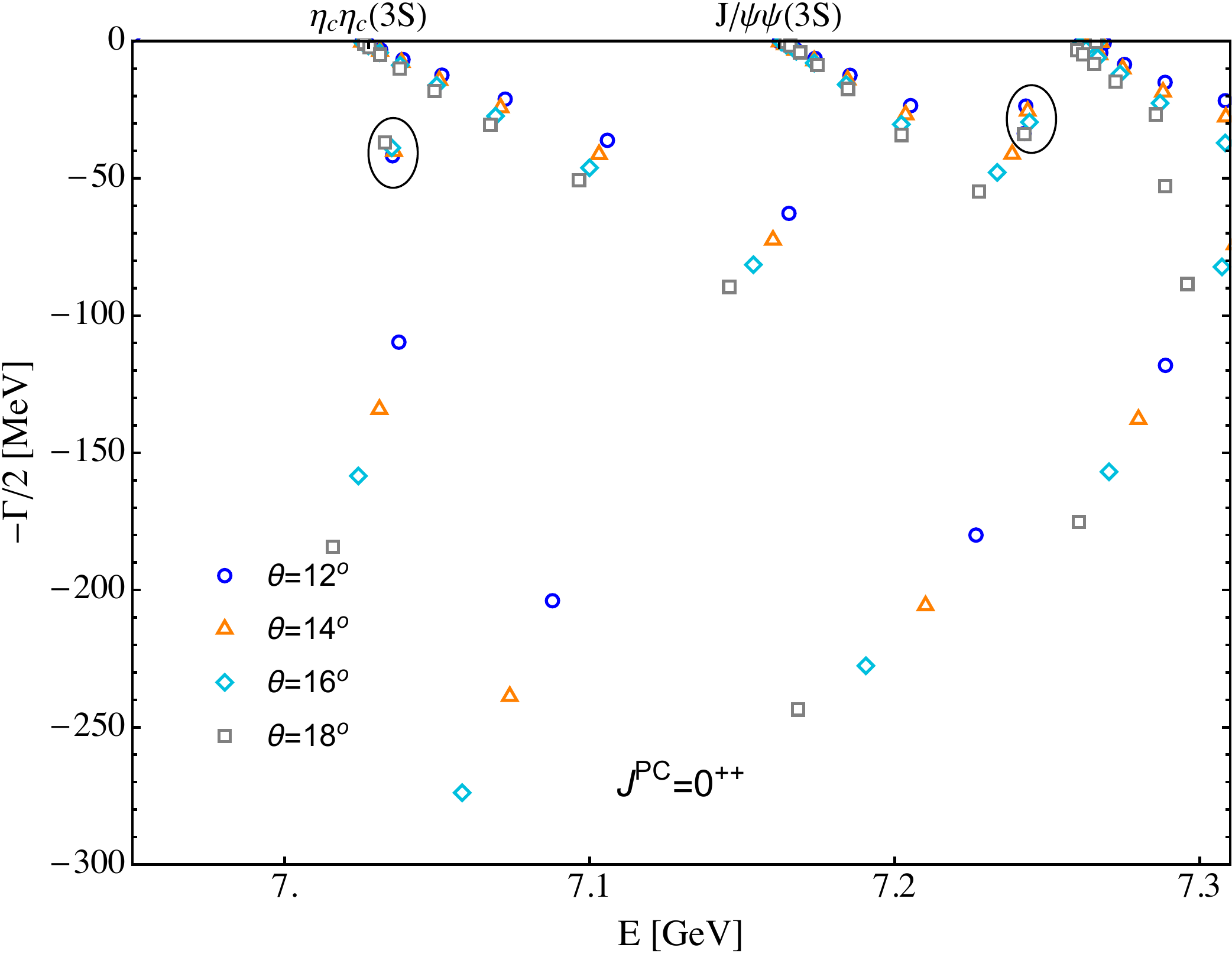}

\end{minipage}%
}%
\centering \caption{(a) The complex energy eigenvalues  of the $0^{++}$ ${cc\bar c\bar c}$ states with varying $\theta$  in the complex scaling method.  (b) The selected view of Fig.~\ref{fig:0++cccc} (a) concentrating on the resonances. The dashed lines represent the continuum lines rotating along $2\theta$.} \label{fig:0++cccc}
\end{figure*}

\begin{figure*}
\centering \subfigure[]{
\begin{minipage}[t]{0.48\linewidth}
\centering
\includegraphics[width=1\textwidth]{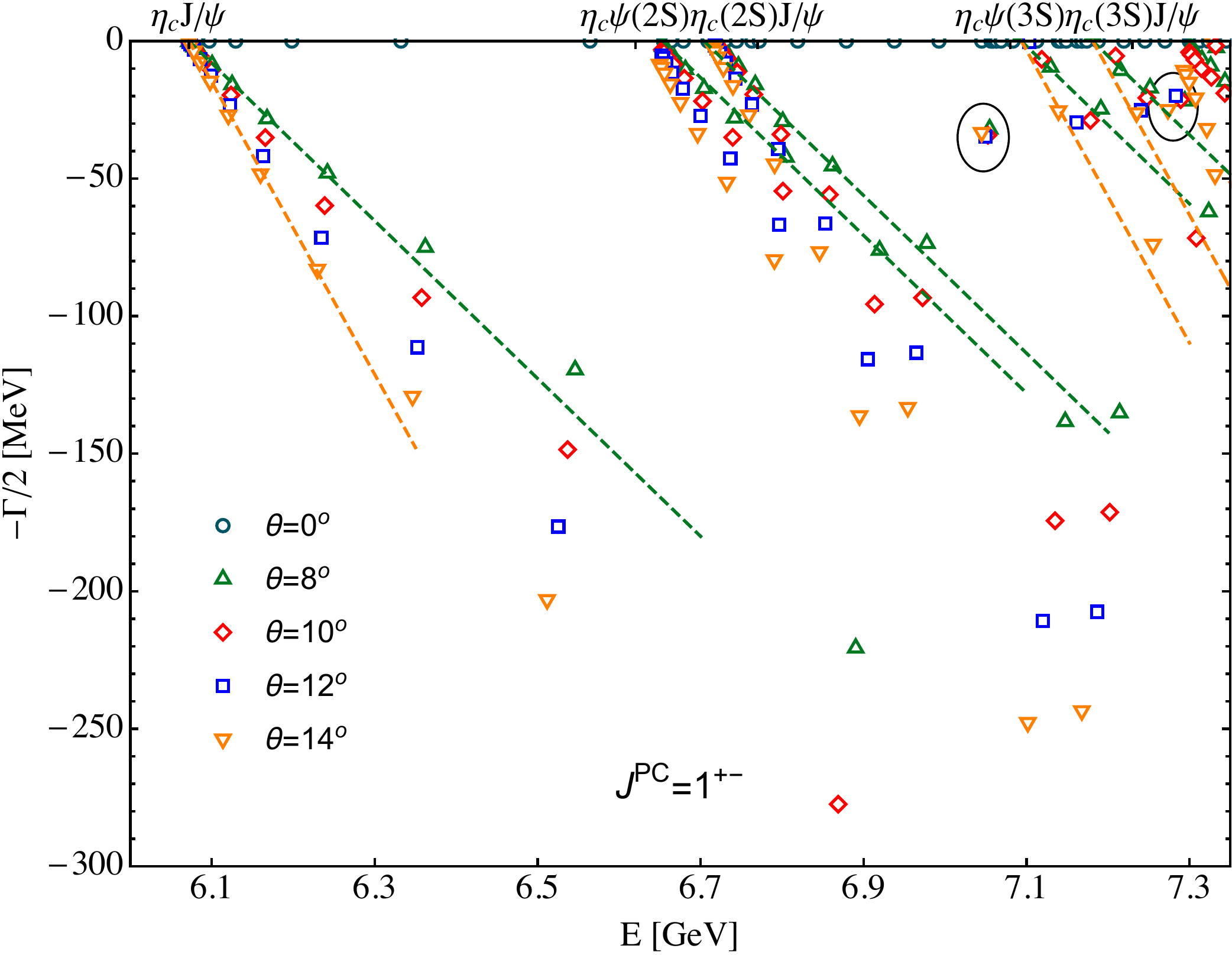}

\end{minipage}%
}%
\subfigure[]{
\begin{minipage}[t]{0.49\linewidth}
\centering
\includegraphics[width=1\textwidth]{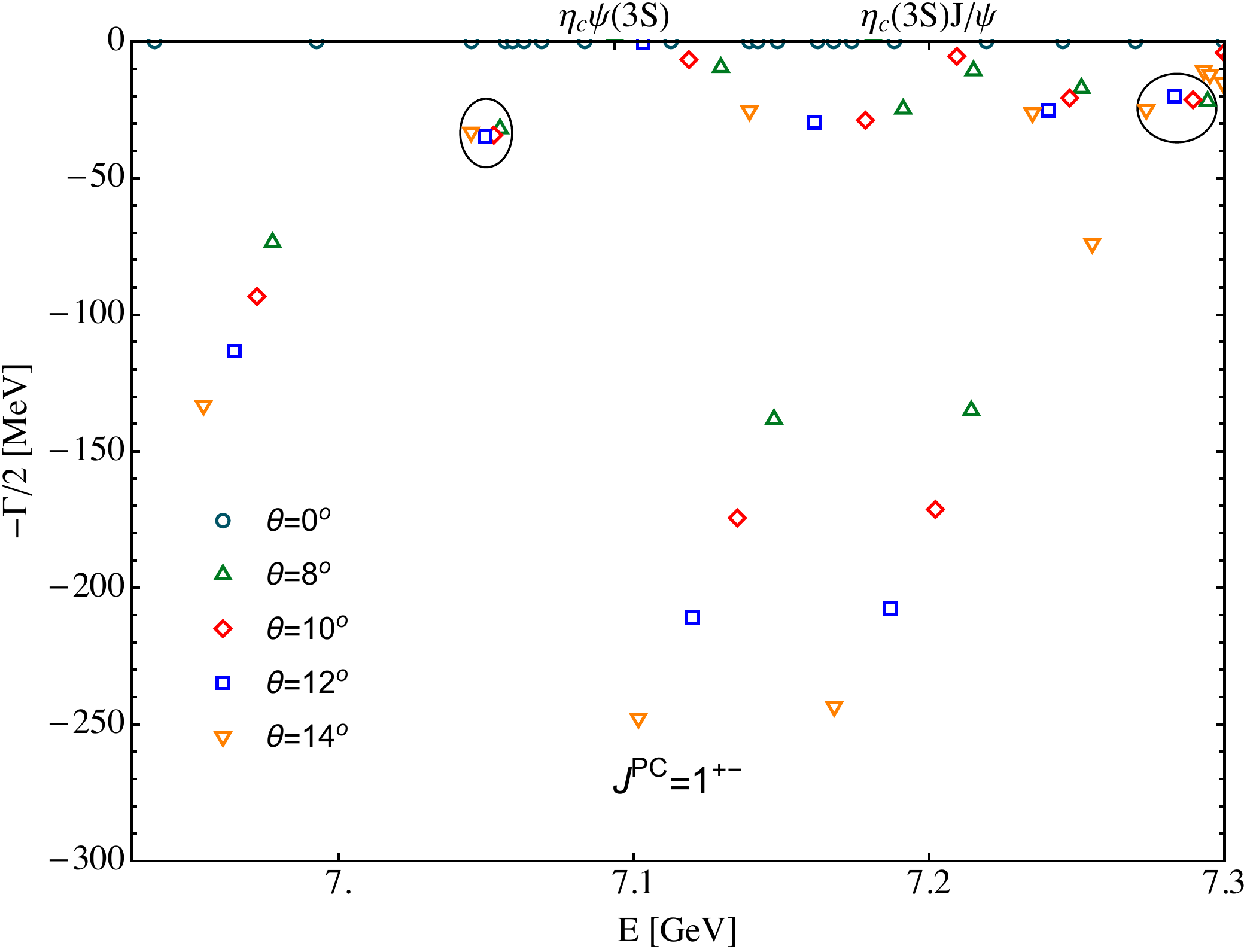}

\end{minipage}%
}%
\centering \caption{(a) The complex energy eigenvalues  of the $1^{+-}$ ${cc\bar c\bar c}$ states with varying $\theta$  in the complex scaling method.  (b) The selected view of Fig.~\ref{fig:1++cccc} (a) concentrating on the resonances. The dashed lines represent the continuum lines rotating along $2\theta$.} \label{fig:1++cccc}
\end{figure*}

\begin{figure*}
\centering \subfigure[]{
\begin{minipage}[t]{0.48\linewidth}
\centering
\includegraphics[width=1\textwidth]{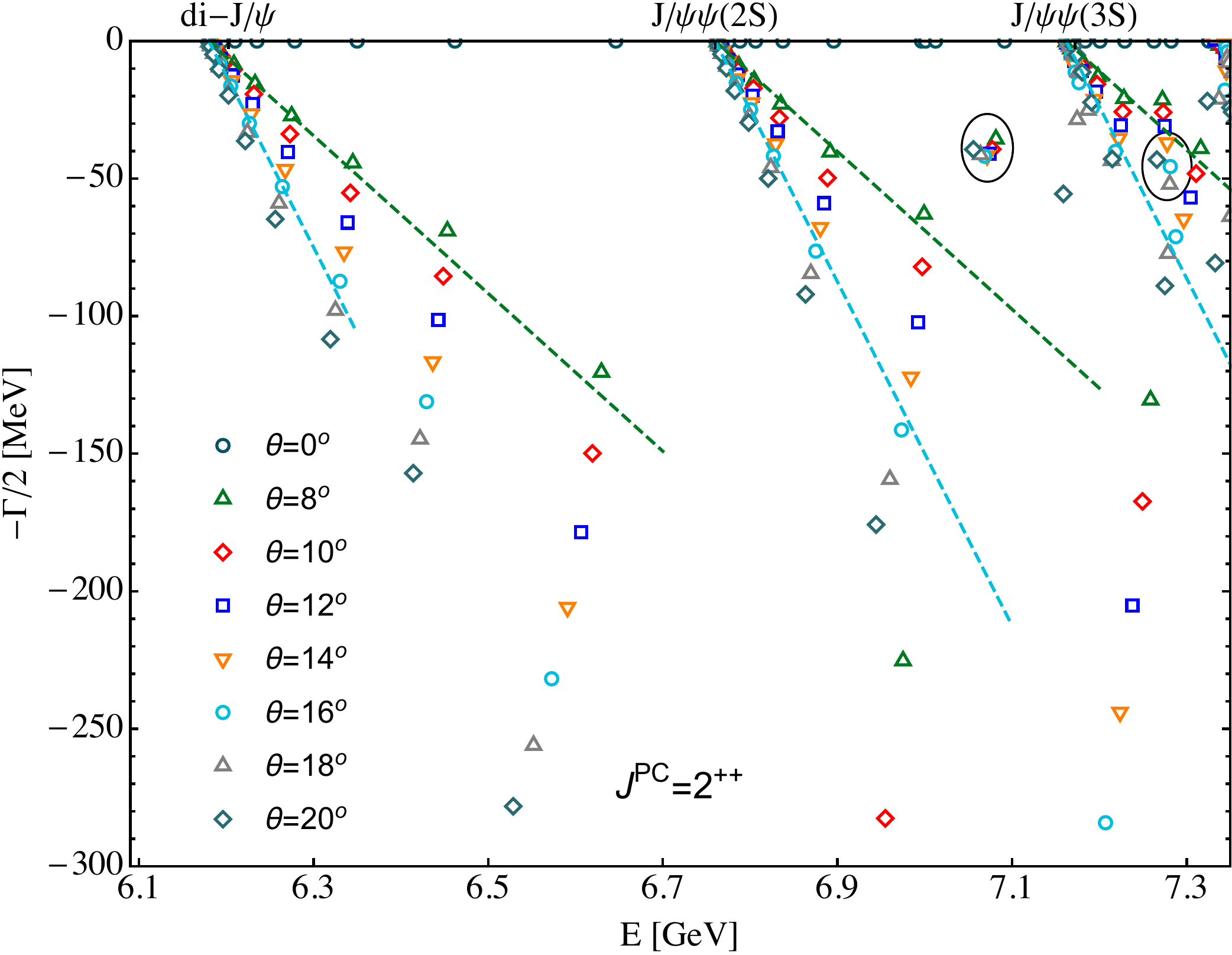}

\end{minipage}%
}%
\subfigure[]{
\begin{minipage}[t]{0.48\linewidth}
\centering
\includegraphics[width=1\textwidth]{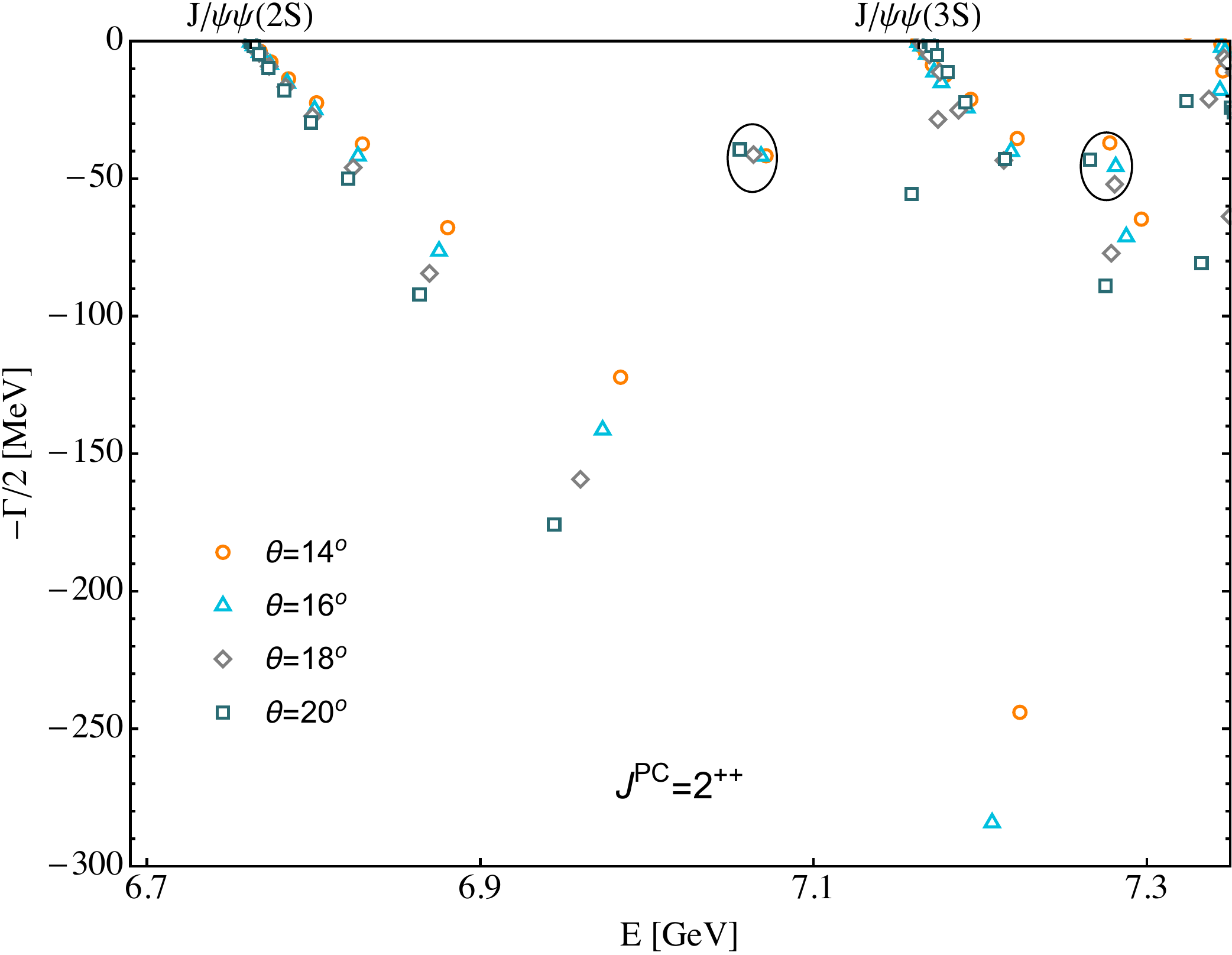}

\end{minipage}%
}%
\centering \caption{(a) The complex energy eigenvalues  of the $2^{++}$ ${cc\bar c\bar c}$ states with varying $\theta$  in the complex scaling method.  (b) The selected view of Fig.~\ref{fig:2++cccc} (a) concentrating on the resonances. The dashed lines represent the continuum lines rotating along $2\theta$.}  \label{fig:2++cccc}
\end{figure*}

To obtain the above resonant states, we have adopted the complex scaling method with the angle $\theta$ ranging from $0$ to $20$ degree. When $\theta=0$, the results correspond to those obtained in real space. The  $\theta$ should be large enough to include the resonance, but not too large, which may lead to strongly oscillating of the asymptotes, especially for the excited states.  The complex eigenvalues  of the $0^{++}$, $1^{+-}$ and $2^{++}$ tetraquark  states with varying $\theta$ are shown in Fig.~\ref{fig:0++cccc}, Fig.~\ref{fig:1++cccc}, and Fig.~\ref{fig:2++cccc}, respectively. As illustrated in Figs.~\ref{fig:0++cccc}-\ref{fig:2++cccc}, we obtain many eigenstates with $\theta=0$. With the complex scaling, most of the states are rotating along the continuum lines, which correspond to the scattering states dicharmonium states as well as their radial excitations.  At the origin of the continuum lines on the real axis with $\theta=0$, we obtained the mass thresholds of the scattering states and summarized the results in Table \ref{tab:scattering states}. As shown in the table, our results of the meson-meson scattering states are consistent with the experimental values up to tens of MeV. With the scattering states as a benchmark, we obtain the masses of the possible resonance and compare those with the experimental values in the following. Since the errors of the scattering channels are tens of MeV, we expect the errors for the tetraquark states to be in the same order.

For the $0^{++}$ $cc\bar c \bar c$ system, we include four possible color-spin configurations as shown in Eqs. \eqref{eq:01} - \eqref{eq:04}. As illustrated in Fig. \ref{fig:0++cccc},  most of the states are rotating along the continuum lines, which correspond to the scattering states  di-$\eta_c$ and di-$J/\psi$ as well as their radial excitations.

\begin{table*}
 \renewcommand\arraystretch{1.5}
\caption{The masses (in the unit of MeV) of the low-lying scattering states obtained with the BGS quark model (BGS)~ \cite{Barnes:2005pb} compared with the experimental values (EXP) \cite{Zyla:2020zbs}.  The script ``--" represents that the mass of the charmonium pair involving $\eta_c(3S)$ is unknown in experiments.}
\label{tab:scattering states}
  \centering
 \setlength{\tabcolsep}{2.3mm}
\begin{tabular}{c|ccccccc}
\toprule[1pt] 
\multirow{3}{*}{$0^{++}$} &  & $\eta_{c}\eta_{c}$ & $J/\psi J/\psi$ & $\ensuremath{\eta_{c}\eta_{c}(2\mathrm{S})}$ & $\ensuremath{\mathrm{J}/\psi\psi^{\prime}}$ & $\ensuremath{\eta_{c}\eta_{c}(3\mathrm{S})}$ & $\ensuremath{\mathrm{J}/\psi\psi(3\mathrm{S})}$\tabularnewline

& BGS  & $5963.7$ & $6180.4$ & $6612.4$ & $6762.2$ & $7025.3$ & $7162.1$\tabularnewline
 & EXP & 5967.8 & 6193.8 & 6621.4  & 6783.0  & -- & 7135.9\tabularnewline
 
\midrule[1pt] 
\multirow{3}{*}{$1^{+-}$} &  & $\eta_{c}J/\psi$ & $\eta_{c}\psi(2S)$ & $\eta_{c}(2S)J/\psi$ & $\eta_{c}\psi(3S)$ & $\eta_{c}(3S)J/\psi$ & \tabularnewline
% & BGS & 6102.2 & 6691.3 & 6704.9 & 7044.3 & 7110.3 & \tabularnewline
  & BGS & 6072.1 & $6653.6$ & $6720.0$ & $7093.6$ & $7181.1$ & \tabularnewline

 & EXP & 6080.8 & 6670.0 & 6734.4 & 7022.9 & -- & \tabularnewline

\midrule[1pt] 
\multirow{4}{*}{$2^{++}$} &  &  & $J/\psi J/\psi$ & $\ensuremath{J/\psi\psi^{\prime}}$ & $\ensuremath{J/\psi\psi}(3S)$ &  & \tabularnewline
 & BGS &   & 6180.4 & 6762.2 & 7162.1 &  & \tabularnewline

 & EXP &  & 6193.8 & 6783.0 & 7135.9 &  & \tabularnewline
\bottomrule[1pt] 
\end{tabular}
\end{table*}

No bound states are observed in the sector. There are two $cc\bar c\bar c$ resonances in the BGS quark model. The lower resonance is located at  the position with $E=7035.1$ MeV and $\Gamma= 77.8$ MeV. It may decay into the di-$\eta_c$, di-$J/\psi$, $\eta_c\eta_c(2S)$, and $J/\psi \psi(2S)$ channels.  Compared with the  $X(6900)$ in experiments, the decay width of the resonance is consistent with the experimental result considering uncertainty, while the mass is about $100$ MeV higher than the experimental central mass.  

The higher resonance is located at the position  $E=7202.2$ MeV and $\Gamma= 60.6$ MeV. The resonant state is located quite close to the threshold lines with a scaling angle in the $(8, 10)$ degree. The coupled channel effect with the scattering states may modify the location of the resonance. To make it clear, we isolate the subdiagram with $\theta$ ranging from a larger degree as illustrated in Fig. \ref{fig:0++cccc} (b). This  state may decay into the  di-$\eta_c$, di-$J\psi$, $\eta_c\eta_c(2S)$,  $J/\psi \psi(2S)$, $\eta_c\eta_c(3S)$,  $J/\psi \psi(3S)$ channels. The mass and width of the higher resonance are consistent with the $X(7200)$, which was observed in  di-$J/\psi$ and $J/\psi \psi(2S)$ channels by CMS~\cite{CMS} and ATLAS~\cite{ATLAS} collaborations, respectively.

For the $1^{+-}$  $cc\bar c \bar c$ system, there are only two possible color-spin configurations as illustrated in Eqs. \eqref{eq:11}-\eqref{eq:12}. As shown in Fig. \ref{fig:1++cccc}, most of the eigenstates obtained at $\theta=0$ degree fall along the continuum cuts corresponding to   $\eta_cJ/\psi$ and their radial excitations. Up to $7.3$ GeV, two resonances survived with the energies and widths as ($E=7049.6$ MeV, $\Gamma=69.4$ MeV) and ($E=7273.5$ MeV, $\Gamma=49.8$ MeV), receptively. They  will both decay into $\eta_cJ/\psi(2S)$, $\eta_c\psi(2S)$, $\eta_c(2S)J/\psi$ channels.  The higher one will also decay into the $\eta_{c}\psi(3S)$ ($7093.6$ MeV) and $\eta_{c}(3S)J/\psi$  ($7181.1$ MeV) channels in BGS quark model. 

For $2^{++}$ $cc\bar c \bar c$ tetraquarks, the possible continuum states are di-$J/\psi$ and the radial excitations.  As shown in Fig. \ref{fig:2++cccc}, two resonances are obtained at the position ($E=7068.5 $, $\Gamma= 83.6 $ MeV) and ($E=7281.3$ MeV, $\Gamma= 91.2$ MeV), respectively. To make the signal of the higher resonance clear, we isolate the subdiagram with $\theta$ ranging from a larger degree as illustrated in Fig. \ref{fig:2++cccc} (b). Given the large enough phase space, the $2^{++}$ resonances will decay into the di-$J/\psi$ and the radially exciting channels. The $0^{++}$ and $2^{++}$ resonances have similar masses and decay widths. Similar to the $0^{++}$ resonances, the two $2^{++}$ resonances may also be the candidates for the $X(6900)$ with $100$  MeV larger mass and $X(7200)$ in experiments. 

So far, we have calculated the $S$-wave $cc\bar c\bar c$ states and did not include the P-wave $cc\bar c\bar c $ states and the scattering states with the orbital excitations. No resonances are found in the energy region $(6.2,6.8)$ GeV, where there may be a $X(6200)$ and a $X(6600)$  as reported by the experiments \cite{CMS, ATLAS}. If the two states exist, there might be two reasons for their absences in the present work. On one hand, if the lower resonance has a very large decay width as reported by the LHCb collaboration~\cite{Aaij:2020fnh}, it may be hard to be studied in the complex scaling method. Such a wide resonance asymptote will oscillate very strongly in the complex plane.  It is difficult to find an ideal set of Gaussian parameters to describe the wave function well. 

Another origin may be the quark model. The quark model has been fixed by the charmonium spectrum, which is suitable for the compact $c\bar c$ system. The flux-tube confinement potential keeps rising linearly with the larger relative radius of two constituent (anti)quarks, which should have terminated at long range and broken into a $q\bar q$ pair. However, such an effect is not considered in the present calculation. In the present quark model, the linear confinement potential keeps increasing with the relative radius. In our model, we include both the diquark-antiquark, the meson-meson configurations, and their coupled channel effect in the  Hamiltonian. The coupling of the resonances and the scattering states may be overestimated when the quark model is directly adopted to describe the quark-quark potential in the coupled channel case.  Due to the strong coupling, some resonance may not survive and act as the continuum states. Moreover, the mass and width of the survived meson will also be affected, which may lead to a worse phenomenology. Besides,  the three or four-body confinement mechanism might exist in the multiquark system. Thus, the improved quark model is expected to study the multiquark system, and the $cc\bar c \bar c$ experimental data may be used to examine the improved confinement mechanism.

\section{Summary} \label{sec3}
In this work, we have calculated the mass spectra of $S$-wave fully charmed tetraquark  $cc\bar c \bar c$ states. Here, we adopt the BGS quark model which is fixed by the charmonium spectrum. The color-Coulomb, the confinement, and the hyperfine interactions are considered for the quark-quark potential. The complex scaling method is used to identify the bound and resonant states from the scattering states. No bound states are obtained. For the $cc\bar c \bar c$ resonances, we found six states, two for $0^{++}$, two for $1^{+-}$,  and two  for $2^{++}$, respectively as  summarized in Table \ref{tab:ccccBGS}.

The lower $0^{++}$ and $2^{++}$ resonances can both decay into the di-$J/\psi$ and the $J/\psi\psi(2S)$ channels. They are located at $7035.1$ and $7202.2$  MeV, respectively, which are around $100$ MeV higher than the experimental $X(6900)$. The higher  $0^{++}$ and $2^{++}$ resonant states have masses and decay widths that are consistent with the $X(7200)$. They might be good candidates for $X(7200)$. The mass shifts between the first and second $S$-wave resonances are about $100$ MeV smaller than that between the $X(6900)$ and $X(7200)$ in experiments. In the quark model, we do not have any free parameters. If both the $X(6900)$ and $X(7200)$ are the $S$-wave tetraquark states,  the inconsistent mass shift may indicate that the quark model should be modified.  Moreover,   we do not find the lower resonances in the mass region $(6.2,6.8)$ GeV with the current quark model.  Our quark model is taken from the compact charmonium spectrum and is extended for the coupling of the diquark-antidiquark and the scattering states. Such a model has not been well justified nor confirmed for the multiquark resonant states.
It may be necessary to modify the model for multiquark systems so that it includes many-body forces and a new confinement mechanism.

\begin{acknowledgements}
We are grateful for the helpful discussions with Prof. Shi-Lin Zhu, Prof. Emiko Hiyama,  Dr. Qian Wu, and Dr. Lu Meng. G.J.~Wang is 
supported by JSPS KAKENHI (No. 20F20026). Qi Meng is supported by the National Natural Science Foundation of China (Grant No. 11822503). M.~Oka is supported in part by the
JSPS KAKENHI (Nos.~19H05159, 20K03959, and 21H00132). 
\end{acknowledgements}

\bibliography{ref}

\end{document}